# Defect-induced exchange bias in a single SrRuO$_3$ layer

Changan Wang,[1,2,3] Chao Chen,[4] Ching-Hao Chang[*,5] Hsu-Sheng Tsai,[1] Parul Pandey,[1] Chi Xu,[1] Roman Böttger,[1] Deyang Chen,[4,6,7] Yu-Jia Zeng[*,3] Xingsen Gao,[4] Manfred Helm,[1,2] and Shengqiang Zhou[*,1]

1) Helmholtz-Zentrum Dresden-Rossendorf, Institute of Ion Beam Physics and Materials Research, Bautzner Landstr. 400, 01328 Dresden, Germany
2) Technische Universität Dresden, D-01062 Dresden, Germany
3) Shenzhen Key Laboratory of Laser Engineering, College of Optoelectronic Engineering, Shenzhen University, 518060 Shenzhen, China
4) Institute for Advanced Materials and Guangdong Provincial Key Laboratory of Quantum Engineering and Quantum Materials, South China Normal University，Guangzhou 510006, China
5) Leibniz-Institute for Solid State and Materials Research, Helmholtzstrasse 20, 01069 Dresden, Germany
6) National Center for International Research on Green Optoelectronics, South China Normal University, Guangzhou 510006, China
7) MOE International Laboratory for Optical Information Technologies, South China Academy of Advanced Optoelectronics, South China Normal University, Guangzhou 510006, China

---

[*] Author to whom correspondence should be addressed. e-mails: c.h.chang@ifw-dresden.de ; yjzeng@szu.edu.cn;s.zhou@hzdr.de

**ABSTRACT**

Exchange bias stems from the interaction between different magnetic phases and therefore it generally occurs in magnetic multilayers. Here we present a large exchange bias in a single SrRuO$_3$ layer induced by helium ion irradiation. When the fluence increases, the induced defects not only suppress the magnetization and the Curie temperature, but also drive a metal-insulator transition at a low temperature. In particular, a large exchange bias field up to ~ 0.36 T can be created by the irradiation. This large exchange bias is related to the coexistence of different magnetic and structural phases that are introduced by embedded defects. Our work demonstrates that spintronic properties in complex oxides can be created and enhanced by applying ion irradiation.

## 1. INTRODUCTION

Exchange bias (EB), a fundamental magnetic interaction, usually manifests itself by a shift in the magnetic hysteresis loop of a ferromagnet. This effect usually stems from the interfacial exchange coupling between a ferromagnet (FM) and an adjacent antiferromagnet (AFM).[1-2] Such an interfacial coupling has been employed for a rich variety of technological applications, including spin-valve device, sensors and magnetic random access memory (MRAM), etc.[3-6] Although EB has been observed in various configurations, such as core/shell structures, epitaxial bilayers and double superlattices, its appearance commonly requires a combination of different materials.[7-10] Observation of EB in a single material is not only physically new but also is highly desirable for practical applications. For instance, a large EB has recently been achieved in compensated ferromagnetic Heusler alloys, which brings a new insight into the mechanism of EB.[11]

Perovskite $SrRuO_3$ (SRO) provides an unique platform for studying EB phenomenon because it has striking electrical and magnetic properties.[12-13] It is the only example of FM ordering ($T_c$ ~ 160 K) in conducting 4d transition-metal oxides of an orthorhombic structure with the *Pbnm* space group.[12, 14] Pioneer studies have confirmed that the actual nature of FM in SRO is closely related to itinerant electrons.[15-16] In an ultrathin SRO film, for example, both the weak localization and the metal-insulator transition (MIT) are generated by the strongly lateral confinement of electrons.[17] Furthermore, a weak EB has been reported in ultrathin SRO films due to the formation of an AFM layer near the substrate.[18-19] On the other hand, physical properties of SRO are susceptible to the lattice structure and therefore they can be tuned efficiently by applying strain on SRO films. A strained SRO film has been shown to have lower magnetization than that in strain-free ones.[20] In addition, theoretical studies have predicted a coexistence of both AFM and FM states in SRO due to the distribution of tensile strain.[21-22] In this respect, the SRO film is a good candidate to study the EB effect in a single material via disturbing itinerant electrons and changing the lattice structure.

In this work, we demonstrate that ion irradiation creates EB effect in single SRO films, instead of multilayer or other heterostructures. Ion irradiation introduced defects not only expand the SRO lattice, but also affect the itinerant electrons. Therefore, the magnetic phases and properties of SRO are modified by irradiation. As the fluence increases, the defect density increases to boost the resistivity, to reduce the magnetization, and to decrease the Curie temperature. MIT is also observed due to the defect-induced weak localization. In particular, a large EB field reaching 0.36 T at 5 K is generated in the irradiated SRO film with a He fluence of $5\times10^{15}$ He/cm$^2$. These results demonstrate that complex oxides can be used for developing novel spintronic devices via applying ion irradiation.

## 2. EXPERIMENTAL SECTION

50 nm SRO thin films were epitaxially grown on SrTiO$_3$ (001) substrates by pulsed laser deposition (PLD) at 680 ºC with an oxygen pressure of 0.1 mbar. After the deposition, Helium (He) ion irradiation was carried out at an energy of 5 keV with a fluence from $1\times10^{15}$ to $5\times10^{15}$ He/cm$^2$. The ion beam was scanned over the sample area to obtain lateral homogeneous irradiation. The samples were referred as $1\times10^{15}$, $2.5\times10^{15}$ and $5\times10^{15}$, respectively. The Stopping and Range of Ions in Matter (SRIM) simulations indicated that He-ions produced a relatively uniform lattice displacement through the SRO thin films.[23] The calculated displacement per atom (DPA) is 2.9%, 7.2% and 14.4% for samples $1\times10^{15}$, $2.5\times10^{15}$ and $5\times10^{15}$, respectively. The structural characterizations were performed by X-ray diffraction (XRD) (PANalytical X'Pert PRO diffractometer) using Cu Kα radiation. The temperature dependence of the resistance of the films was measured using the van der Pauw method in a Lakeshore Hall measurement system. Magnetic properties were measured with a superconducting quantum interference device (Quantum Design, SQUID-VSM) magnetometer.

## 3. RESULTS AND DISCUSSION

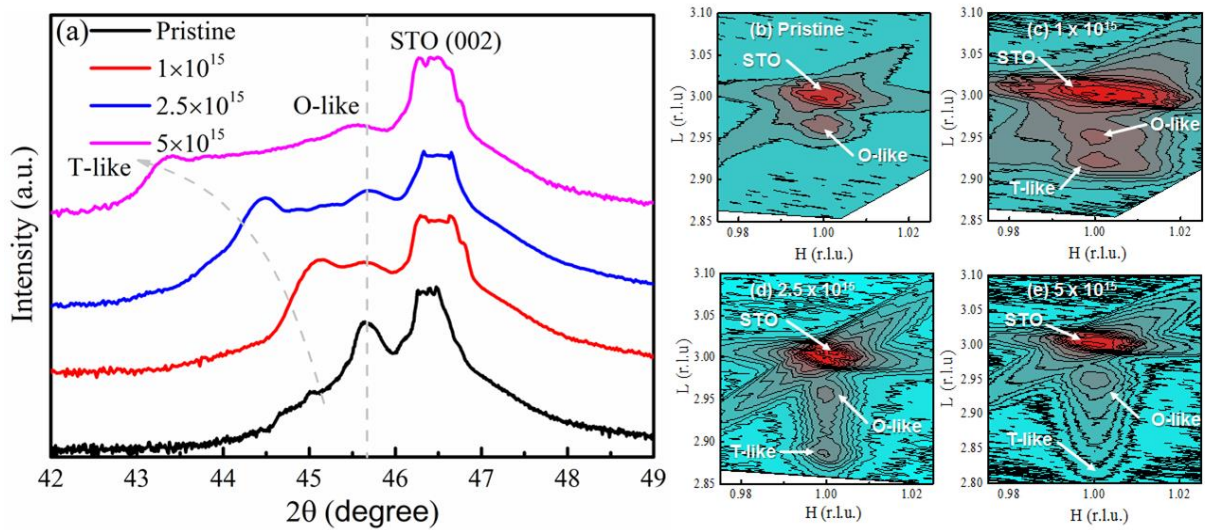

**Figure 1.** (a) θ-2θ scans around the (002) reflection of SRO thin films on STO substrates under different He fluences. (b)-(e) Corresponding reciprocal-space mappings around the (103) reflections of SRO films.

Figure 1a shows XRD θ-2θ scans of the SRO films around the (002) reflection of the STO substrate before and after irradiation. The pristine SRO film on the STO substrate ($a_{sub}$ = 3.905Å) is strained compressively. A small signal belonging to the tetragonal-like (T-like) phase is observed to coexist with that of the orthorhombic-like (O-like) phase. Such a mixed phase is probably due to compressive strain from the substrate that stabilizes the T-like phase.[24-25] After He ion irradiation, the intensity of O-like peak becomes weaker and its position remains almost the same for different fluences. The T-like peak, however, not only increases in intensity but also shifts towards lower angles as the fluence increases. These results indicate that the average out-of-plane lattice constant is expanded due to the He implantation and thus results in a transition from O-like phase to T-like phase.[26] The reciprocal-space mappings (RSMs) collected around the (103) peak of SRO films under different He fluences are shown in Figure 1b-e. It can be seen that the pristine film is epitaxially grown on the STO substrate, since the in-plane lattice parameters of SRO films are identical to that of the substrate (see Figure 1b). When the fluence increases, diffraction peaks of O-like and T-like phase are found to share the same H values as that of the substrate, which suggests that the in-plane lattice constants of SRO are still epitaxially locked to that of the STO substrate even after the irradiation. Differently, the T-like peak shifts towards low L values, which is consistent with the θ-2θ diffractogram in Figure 1a. Both measurements demonstrate that the irradiation induces the expansion of the out-of-plane lattice constant and therefore triggers a transition from an O-like phase to a T-like one.

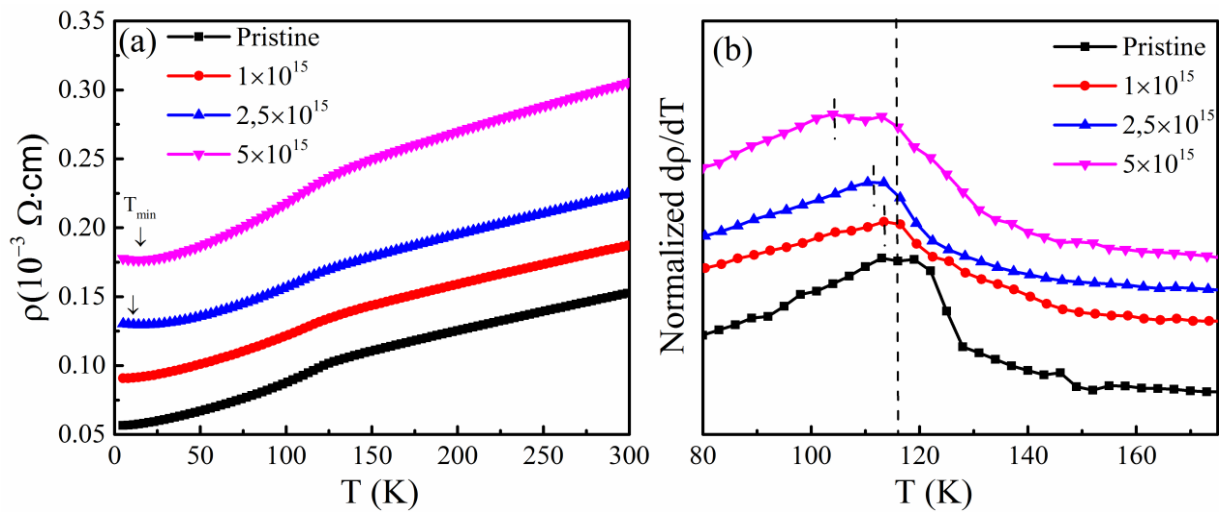

**Figure 2.** (a) Temperature dependence of resistivity of the SRO films with various He-ion fluence. Arrows mark the temperature where upturns in the resistivity occur. (b) Corresponding dρ/dT versus T curves.

The temperature dependence of resistivity is presented in Figure 2a for SRO films with different He fluences. Although the pristine SRO contains two phases, the O-like phase and T-like phase all show similar temperature dependent resistivity.[27] The resistivity increases continuously as the fluence of ions increases. This upsurge can be understood by the enhanced disorder due to the increased fluence. Upon closer inspection, the pristine and the ion-irradiated ($1\times10^{15}$ He/cm$^2$) films remain metallic as the temperature is decreased to 5 K. The SRO films implanted with the higher fluences of $2.5\times10^{15}$ and $5\times10^{15}$ He/cm$^2$, however, become insulating below a certain temperature, $T_{min}$, at ~ 14 K and 16 K, respectively (see upturns in Figure 2a). The MIT is probably due to the disorder-induced weak localization.[28-29] The kink around 120 K in the resistivity in Figure 2a corresponds to a phase transition from paramagnetism to ferromagnetism near the Curie temperature ($T_C$). The magnetic phase transition couples to the electric transport in ferromagnetic metals that suppresses spin dependent scattering and then influences the resistivity of the SRO.[28] To clearly distinguish this phase transition, the derivative of resistivity against temperature is plotted in Figure 2b, in which the transition temperature, $T_C$ can be easily identified. This transition temperature decreases from 116 K to 107 K, as the fluence increases from 0 He/cm$^2$ (Pristine) to $5\times10^{15}$ He/cm$^2$. The film implanted with the highest fluence of $5\times10^{15}$ He/cm$^2$ shows two peaks, which originate from O-like and T-like phases. It has been reported that the transition temperature of SRO is highly sensitive to the cell distortion.[27] As shown in Figure 1a, the modification of the lattice structure of T-like phase is more suspect to ion irradiation and the perpendicular lattice spacing is extended after irradiation. Therefore, the peaks are separated remarkably at $5\times10^{15}$ sample in Figure 2b. Other samples do not exhibit this phenomenon due to the overlap of peaks.[27] The magnetic behavior of SRO films is known to be very sensitive to the cell distortion or the $RuO_6$ octahedral tilting.[15, 28, 30] According to Figure 1, in our films defects induced strain results in structural distortion by expanding the out-of-plane lattice parameter *c*, which has also been observed in He irradiated $La_{0.7}SrMnO_3$ films.[26, 31] Therefore, the decrease of $T_C$ in our SRO films can be attributed to the weakened ferromagnetic interaction due to the modulation of Ru-O-Ru bonds as *c* is expanded.

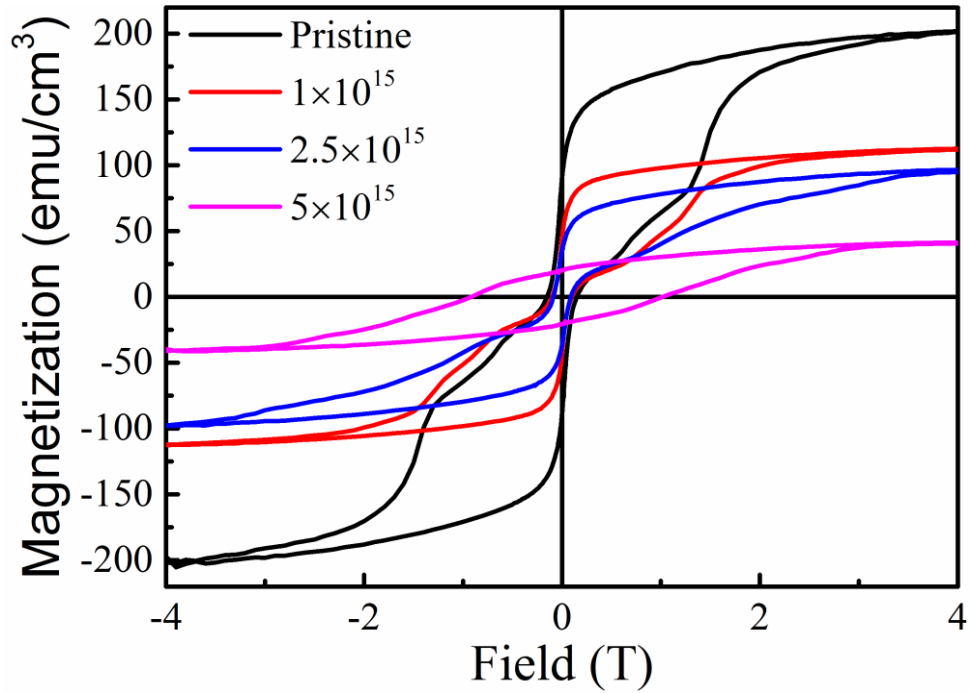

**Figure 3.** Magnetic-hysteresis (*M-H*) loops at 5 K for the SRO films irradiated with different He ion fluences.

Figure 3 presents magnetic-hysteresis (*M-H*) loops of SRO films with different He-ion fluences at 5 K. All the samples show hysteresis behavior, indicating typical ferromagnetism in SRO films.[20, 32] In the hysteresis curves, the magnetization shows jumps at different field, which is generally attributed to the Barkhausen jump.[33] These jumps are commonly caused by the irreversible motion of the domain walls between two regions of opposite magnetizing forces.[32-34] When an external magnetic field is applied on these domain walls, the magnetization of the materials changes in a series of discontinuous steps and then causes "jumps" in the magnetic field. Besides, the different crystallography variants in SRO might also contribute to the jumps.[35] Most remarkably, the magnetization in these films decreases with the increase of the irradiation fluence. It can be attributed to the modulation of lattice structure due to the He ion irradiation. The larger structure distortion induced by the He ion irradiation leads to a change in the spin-spin coupling, which results in changes in the spin-split electronic band structure and the exchange energy among spins.[32, 36-37] In addition to the variation of spin-spin coupling, the change of magnetic anisotropy as the component of T-like phase increases can also contribute to the decrease of magnetization.

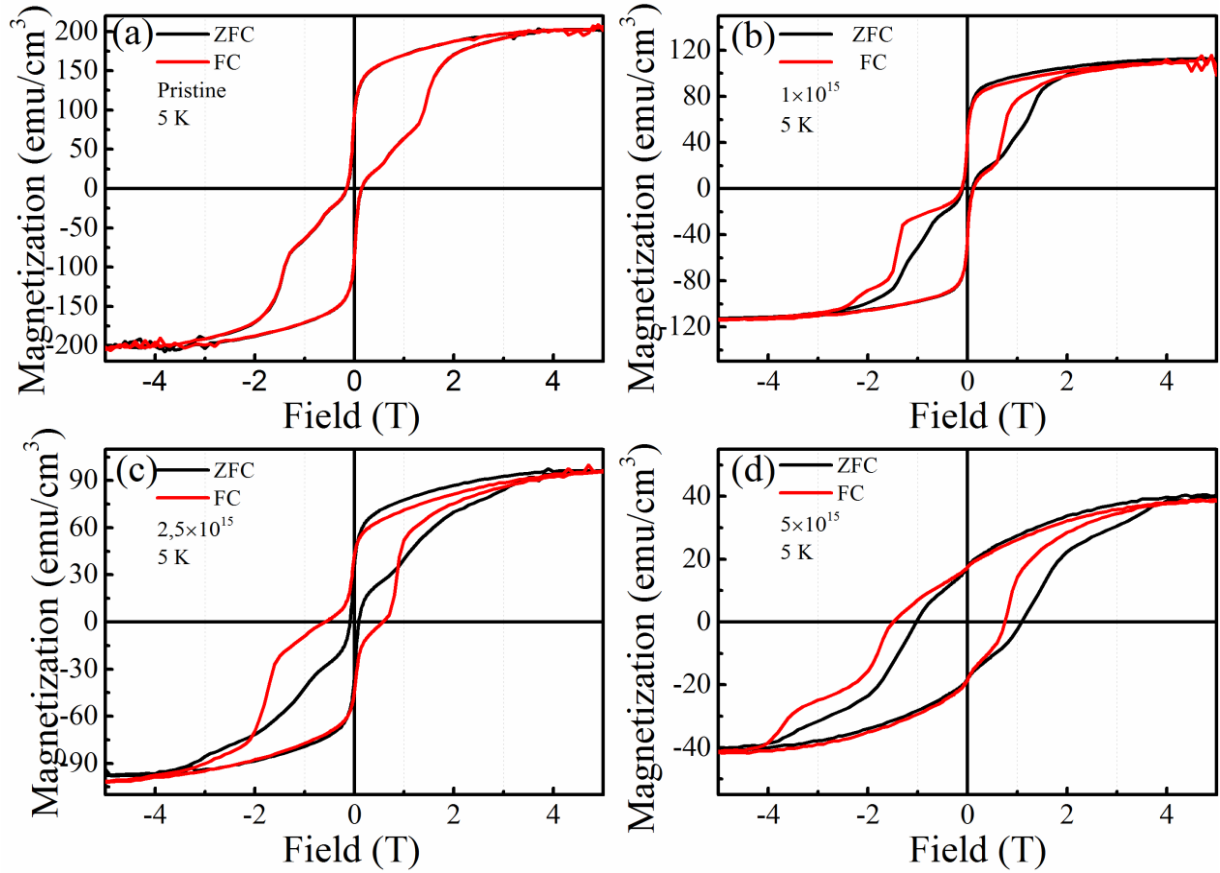

**Figure 4.** (a)-(d) ZFC/FC magnetic-hysteresis (*M-H*) loops for the SRO films of different He-ion fluence measured at 5 K.

Having established that material structures and magnetic phases are susceptible to ion irradiation, we next study whether the EB effect can be introduced in a single SRO layer. Figure 4 presents the *M-H* loops for the SRO films with different He-ion fluences. The results are measured at 5 K after sample being cooled from 300 K with and without applied magnetic field of 1 T. For the pristine sample both of the zero-field cooling (ZFC) *M-H* curve and field cooling (FC) *M-H* curve exhibit symmetric hysteresis loops centered at the origin. However, for the FC *M-H* curves of the films with irradiation, the loops are nonsymmetric with respect to the origin and are biased toward the negative field, which clearly suggests the presence of the EB effect.[1] This effect is strongly influenced by the magnitude of the irradiation fluence, where a larger fluence leads to a larger EB field. An EB field ($H_{EB}$) is defined as $H_{EB}=|(H_1+H_2)/2|$, where $H_1$ is the field where the ferromagnetic moment becomes negative and $H_2$ is the field where the moment changes from negative to positive. A maximum value up to ~ 0.36 T at 5 K is obtained in the SRO film implanted with the highest fluence.

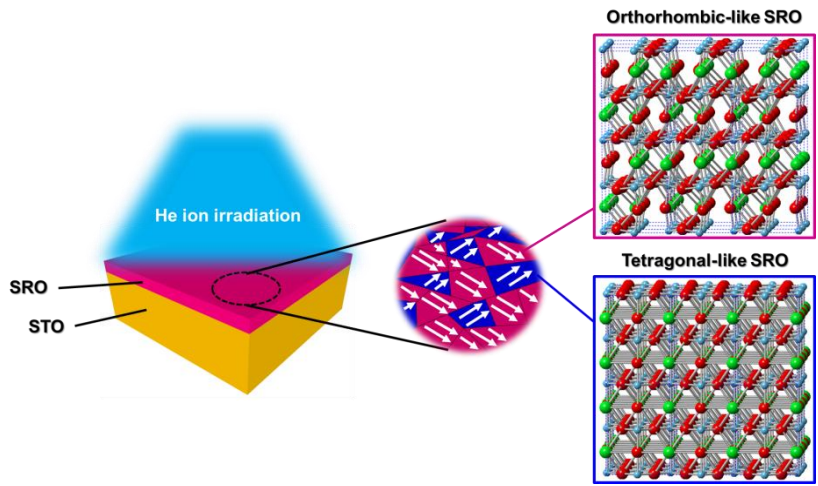

**Figure 5.** Schematic illustration of the magnetic phases with different easy axes in an irradiated SRO film.

The observed EB effect probably stems from the interaction between different magnetic phases that are susceptible to ion irradiation. A simplified illustration of the spins of different phases for the irradiated SRO films is shown in Figure 5. In FM/AFM heterostructures, EB effect has been observed, which is attributed to pinned/biased moments and uncompensated spins at the interface.[2, 7] Whereas in the FM/FM systems, this phenomenon is also observed, which is explained by the strong interaction between the microscopic interface domain structure and the uniaxial magnetic anisotropy (MA).[8,38] In our case, it has been reported that the T-like phase has out-of-plane uniaxial magnetic anisotropy (MA), whereas the O-like phase exhibits in-plane uniaxial MA.[39-40] The significant difference of MA in these two phases may result in the observed EB in our system. In SRO films, the signal of T-like phase is weak in the pristine sample and then becomes significant as the fluence increases (see Figure 1a). This feature finally increases the strength of the interaction between T-like and O-like phases, leading to a large EB in the SRO film irradiated with the fluence of $5\times10^{15}$ He/cm$^2$. In addition to the MA-induced EB, disorder induced by the ion irradiation may drive an antiferromagnetic phase in SRO, which can couple with the primary ferromagnetic phase to contribute to the EB.[19] Therefore, our results demonstrate that the large EB effect is a consequence of artificial defects in SRO films, which can be controllably introduced by He ion irradiation. This study paves a way to develop novel spintronic materials by applying ion irradiation. Ion irradiation is indeed a well-developed technique and has been widely used for modern chip technologies.

## 4. CONCLUSIONS

In summary, we show a pronounced exchange bias in the single ferromagnetic SRO film by helium irradiation. When the fluence increases, the induced defects result in the enhancement of resistivity and the reduction of magnetization as well as the decrease of Curie temperature. The metal-insulator transition occurs in irradiated samples due to the disorder-induced weak localization. The obtained exchange bias field is up to ~ 0.36 T, which is larger than those reported in other systems.[18, 22] These emergent physical phenomena are attributed to defect-induced disorder and structural distortion. This work shows a framework for designing spintronic properties of complex oxides by applying ion irradiation.

**ACKNOWLEDGMENTS**

C. A. Wang thanks China Scholarship Council (File No. 201606750007) for financial supports. C.-H.C. and P. Pandey acknowledge financial support from the German Research Foundation (Deutsche Forschungsgemeinschaft, Grant CH 2051/1-1 and ZH 225/6-1, respectively). This work was also supported by the National Natural Science Foundation of China (Nos. 51502178，11704130) and the Ministry of Science and Technology through Grant (No. 107-2917-I-564-034-).